\documentclass[12pt]{article}
\usepackage{amssymb,diagrams}
\newcommand{\idty}{{\leavevmode{\rm 1\mkern -5.4mu I}}}
\newcommand{\be}{\begin{eqnarray}}
\newcommand{\ee}{\end{eqnarray}}
\newcommand{\A}{{\cal A}}

\renewcommand{\O}{\Omega}

\renewcommand{\d}{{\rm d}}
\newcommand{\bez}{\begin{eqnarray*}}
\newcommand{\eez}{\end{eqnarray*}}
\newcommand{\dl}{{\delta}}
\newcommand{\Dl}{{D_\delta}}
\title{\bf Bi-differential calculi and integrable models}
\date{  }
\author{A. Dimakis \\ Department of Mathematics, University of the Aegean \\
        GR-83200 Karlovasi, Samos, Greece \\ dimakis@aegean.gr
        \\[2ex]
        F. M\"uller-Hoissen \\ Max-Planck-Institut f\"ur Str\"omungsforschung \\
        Bunsenstrasse 10, D-37073 G\"ottingen, Germany \\
        fmuelle@gwdg.de }

\begin{document}
\renewcommand{\theequation} {\arabic{section}.\arabic{equation}}

\maketitle

\begin{abstract}
The existence of an infinite set of conserved currents in completely integrable classical models, including chiral and Toda models as well as the KP and self-dual Yang-Mills equations, is traced back to a simple construction of an infinite chain of closed (respectively, covariantly constant) 1-forms in a (gauged) bi-differential calculus. The latter consists of a differential algebra on which two differential maps act. In a gauged bi-differential calculus these maps are extended to flat covariant derivatives.
\end{abstract}

\section{Introduction}
\setcounter{equation}{0}
Soliton equations and other completely integrable models are distinguished by the existence of an infinite set of conservation laws. In particular, for two-dimensional (principal) chiral or $\sigma$-models an infinite set of nonlocal conserved currents had been found \cite{Lues+Pohl78} and later a simple iterative construction has been presented \cite{BIZZ79}. The latter construction was formulated in terms of differential forms and then generalized to noncommutative differential calculi 
on commutative algebras by the present authors \cite{DMH96,DMH97}, and moreover to differential calculi on noncommutative algebras \cite{DMH98}. As a particular example, this generalization includes the case of the nonlinear Toda lattice and the corresponding nonlocal conserved charges coincide with those which had been obtained earlier in a different way \cite{Heno74}. The question then arose whether also other soliton models, like KdV, fit into this scheme. In this work we present a somewhat radical abstraction of the abovementioned iterative construction of conserved currents for chiral models which indeed applies to many of the known soliton equations and integrable models. It severely deviates, however, from our previous approach \cite{DMH96,DMH97} which made use of a generalized Hodge operator on noncommutative differential calculi. 
Our present approach is based on differential calculi with two differential maps (which are analogues of the exterior derivative of the differential calculus on manifolds). We are thus dealing with {\em bi}-differential calculi, a structure which we introduce in section 2. Section 3 contains as an example a generalization of Plebanski's first heavenly equation \cite{Pleb75,Taka89}. Most interesting examples require a `gauged bi-differential calculus' which we consider in section 4. Several integrable models which fit into this framework are presented in section 5. Some concluding remarks are collected in section 6.

\section{Bi-differential calculi and an iterative construction of closed forms}
\label{sec:bi-differential}
\setcounter{equation}{0}
Let $\A$ be an associative algebra\footnote{We consider algebras over $\mathbb{R}$ or $\mathbb{C}$. A {\em linear} map is then linear over $\mathbb{R}$, respectively $\mathbb{C}$.}. A {\em graded algebra over} $\A$ is a $\mathbb{N}_0$-graded associative algebra $\O(\A) = \bigoplus_{r \geq 0} \O^r(\A)$ where $\O^0(\A) = \A$. Furthermore, we assume that $\O(\A)$ has a unit $\idty$ such that $\idty \, w = w \, \idty = w$ for all  $w \in \O(\A)$. A {\em differential calculus} $(\O(\A),\d)$ over $\A$ consists of a graded algebra $\O(\A)$ over $\A$ and a linear map $ \d \, : \,  \O^r (\A) \rightarrow \O^{r+1}(\A)$ with the properties
\be
       \d^2 &=& 0 \\
       \d (w \, w') &=& (\d w) \, w' + (-1)^r \, w \, \d w'  \label{Leibniz}
\ee
where $w \in \O^r(\A)$ and $w' \in \O (\A)$. The identity $\idty \idty = \idty$ then implies $ \d \idty = 0 $. 
% We also require that $\d$ generates $\O(\A)$ in the sense that 
% $\O^{r+1}(\A) = \A \, (\d \O^r(\A)) \, \A$. 
A triple $(\O(\A),\d , \delta)$ consisting of a graded algebra $\O(\A)$ over $\A$ and two maps 
$\d , \delta \, : \, \O^r (\A) \rightarrow \O^{r+1}(\A)$ with the above properties and
\be
       \delta \, \d + \d \, \delta = 0   \label{d-delta}
\ee
we call a {\em bi-differential calculus}.

\vskip.2cm
\noindent
{\em Example.} The following sets up a (somewhat restricted) framework for constructing bi-differential calculi. All the examples which we encounter in the following sections actually fit into this scheme.
Let $\xi^\mu$, $\mu = 1, \ldots, n$, generate $\O^1(\A)$ as a left $\A$-module. This requires commutation rules for $\xi^\mu$ and the elements of $\A$.
Assuming $\xi^\mu \xi^\nu + \xi^\nu \xi^\mu =0$, products of the $\xi^\mu$ then generate $\O(\A)$ as a left $\A$-module. Let $M_\mu, N_\nu$ be derivations $\A \rightarrow \A$. We define
\be
    \d f = (M_\mu f) \, \xi^\mu \, , \qquad  \delta f = (N_\mu f) \, \xi^\mu 
\ee
and
\be
    \d (f \, \xi^{\mu_1} \cdots \xi^{\mu_r}) = (\d f) \, \xi^{\mu_1} \cdots \xi^{\mu_r} \, , \qquad  
    \delta (f \, \xi^{\mu_1} \cdots \xi^{\mu_r}) = (\delta f) \, \xi^{\mu_1} \cdots \xi^{\mu_r}
\ee
($r=1, \ldots, n$). Then we have
\be
  \d^2 = 0  \quad &\Longleftrightarrow&  \quad [ M_\mu , M_\nu ] =0  \\
  \delta^2 = 0  \quad &\Longleftrightarrow&  \quad [ N_\mu , N_\nu ] =0  \\
  \d \, \delta + \delta \, \d = 0 \quad &\Longleftrightarrow& \quad  
    [ M_\mu , N_\nu ] = [ M_\nu , N_\mu ]  \; .
\ee
Up to this point we did not have to specify the commutation rules between the $\xi^\mu$ and the elements of $\A$. The graded Leibniz rule (\ref{Leibniz}) holds, in particular, if $[\xi^\mu , f]=0$ for all $f \in \A$ and $\mu =1, \ldots, n$.
\hfill                $\diamondsuit$
\vskip.2cm

Let $(\O(\A), \d , \delta)$ be a bi-differential calculus such that, for some $s \geq 1$, the $s$-th cohomology $H^s_\delta (\O(\A))$ is trivial, so that all $\delta$-closed $s$-forms are $\delta$-exact. Suppose there is a (nonvanishing) $\chi^{(0)} \in \O^{s-1}(\A)$ with 
\be
         \delta \chi^{(0)} =0   \; .      \label{del_chi0=0}
\ee
Let us define
\be
     J^{(1)} = \d \chi^{(0)}   \; .
\ee
Then
\be
        \delta J^{(1)} = - \d \delta \chi^{(0)} = 0   
\ee
so that
\be
       J^{(1)} = \delta \chi^{(1)}
\ee
with some $\chi^{(1)} \in \O^{s-1}(\A)$. Now let $J^{(m)}$ be any $s$-form which satisfies
\be
       \delta J^{(m)} = 0 \, , \qquad  J^{(m)} = \d \chi^{(m-1)}   \; .
\ee 
Then
\be
        J^{(m)} = \delta \chi^{(m)}           \label{J=dchi}
\ee
with some $\chi^{(m)} \in \O^{s-1}(\A)$. Hence
\be
        J^{(m+1)} = \d \chi^{(m)}  
\ee
is $\delta$-closed:
\be
  \delta J^{(m+1)} = - \d \delta \chi^{(m)} = - \d J^{(m)} = - \d^2 \chi^{(m-1)} = 0  \; .
\ee
In this way we obtain an infinite tower\footnote{Of course, in certain examples this construction may lead to something trivial. This happens, in particular, if $\d$ and $\delta$ are linearly dependent, so that the bi-differential calculus reduces to a differential calculus. See also the first remark below.}
of $\delta$-closed $s$-forms $J^{(m)}$ and elements $\chi^{(m)} \in \O^{s-1}(\A)$ satisfying
\be
      \delta \chi^{(m+1)} = \d \chi^{(m)}  \label{delta-chi-d-chi}
\ee
(see Fig.~1).

\diagramstyle[PostScript=dvips]
\begin{diagram}[notextflow]
 &         &\chi^{(0)}&         &       &         &\chi^{(1)}&         &       &         &\chi^{(2)}&&& \\
 &\ldTo^\dl&          &\rdTo^\d &       &\ldTo^\dl&          &\rdTo^\d &       &\ldTo^\dl&          &\rdTo^\d&& \\
0&         &          &         &J^{(1)}&         &          &         &J^{(2)}&         &          & &J^{(3)}&\cdots\\
 &         &          &\ldTo^\dl&       &\rdTo^\d &          &\ldTo^\dl&       &\rdTo^\d &          &\ldTo^\dl&&  \\
 &         & 0        &         &       &         & 0        &         &       &         & 0        &&&
\end{diagram}

\vskip.1cm \noindent
\begin{center}
{\bf Fig.~1}  \\
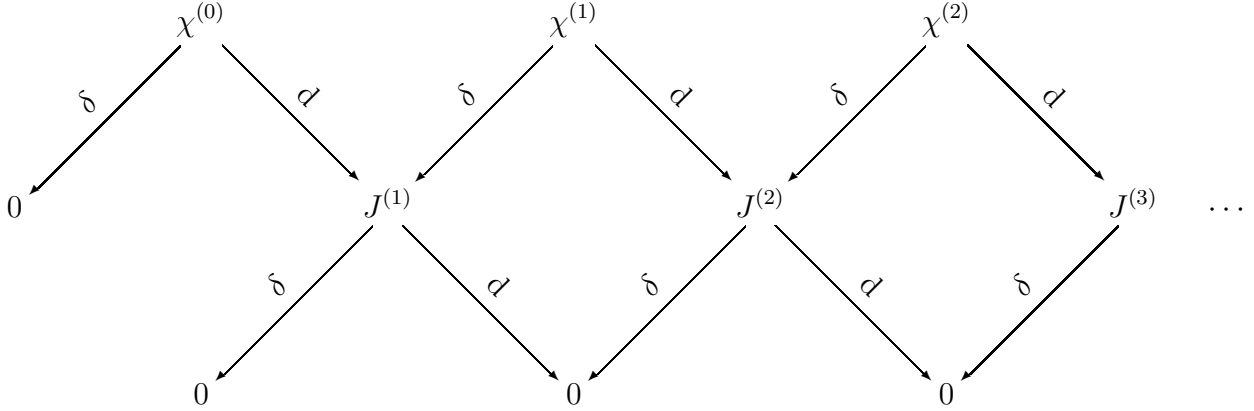
The infinite tower of $\delta$-closed $s$-forms $J^{(m)}$.
\end{center}
\vskip.2cm

Introducing\footnote{In general, we can only expect $\chi$ to exist as a formal power series in $\lambda$.}
\be
    \chi = \sum_{m=0}^\infty \lambda^m \, \chi^{(m)}   \label{chi_sum_sec3}
\ee
with a parameter $\lambda$, the set of equations (\ref{delta-chi-d-chi}) implies
\be
    \delta \chi = \lambda \, \d \, \chi \; .    \label{chi_eq_sec3}
\ee
Conversely, if this equation holds for all $\lambda$, we recover (\ref{delta-chi-d-chi}).

\vskip.2cm
\noindent
{\em Remarks.} \\
(1) Let $\chi^{(0)} \in \O^{s-1}(\A)$, $s>1$, such that $\chi^{(0)} = \delta \alpha$ with some $\alpha \in \O^{s-2}(\A)$. This leads to $J^{(1)} = \d \chi^{(0)} = \d \delta \alpha = -\delta \d \alpha$ and thus $\chi^{(1)} = -\d \alpha$ (up to addition of some $\beta \in \O^{s-1}(\A)$ with $\delta \beta =0$, see the following remark). Hence $J^{(2)} =0$, and the construction of $\delta$-closed $s$-forms breaks down at the level $m=2$.  In order to have a chance, that the iteration procedure produces something nontrivial at arbitrarily high levels $m$, it is therefore necessary that the cohomology $H^{s-1}_\delta (\O(\A))$ is not trivial and the iteration procedure must start with some $\chi^{(0)} \in \O^{s-1}(\A)$ which is $\delta$-closed, but not $\delta$-exact.  \\
(2) $\delta J^{(m)} =0$, $m>0$, determines $\chi^{(m)}$ via (\ref{J=dchi}) only up to addition of some $\chi^{(0)}_m$ with $\delta \chi^{(0)}_m =0$. But $\chi^{(0)}_m$ then plays the same role as $\chi^{(0)}$ ! Hence, this freedom corresponds to a new chain starting at the $m$th level. If there is only a single linearly independent $\chi^{(0)} \in \O^{s-1}(\A)$ with $\delta \chi^{(0)} =0$  (but $\chi^{(0)}$ not $\delta$-exact), this means that $J^{(m)}$ is determined only up to addition of some linear combination of the $J^{(q)}$ with $1 \leq q < m$. In this case, we are loosing nothing by simply ignoring the above freedom in the choice of $\chi^{(m)}$. If there are several linearly independent choices for $\chi^{(0)}$ (with $\delta \chi^{(0)} =0$, but $\chi^{(0)}$ not $\delta$-exact), we have to elaborate the sequences $J^{(m)}$, $m>0$, for all of these choices (respectively, for their general linear combination). Again, the freedom in the choice of $\chi^{(m)}$, $m>0$, then does not lead to anything new. \\
(3) In the definition of a bi-differential calculus we have assumed that both differential maps $\d$ and $\delta$ act on the same grading of $\O(\A)$. The above iteration procedure works, however, as well if they operate on different gradings. Then we have to start with a bi-graded algebra $\O(\A) = \bigoplus_{r \geq 0, s \geq 0} \O^{r,s}(\A)$ with $\O^{0,0}(\A)=\A$, and differential maps $\d \, : \, \O^{r,s}(\A) \rightarrow \O^{r+1,s}(\A)$, $\delta \, : \, \O^{r,s}(\A) \rightarrow \O^{r,s+1}(\A)$ satisfying (\ref{d-delta}). \\       
(4) In classical differential geometry, bi-differential calculi appeared under the name {\em double complex} or {\em bicomplex} (see \cite{FGG76}, for example). In particular, given a differentiable fibre bundle, a splitting of the exterior derivative on the bundle space into vertical and horizontal\footnote{Horizontal with respect to a (local) cross section or flat connection.} parts leads to a bicomplex. In this way, bicomplexes also appeared in the context of symmetries and conservation laws of Euler-Lagrange systems (see \cite{Vino78}, for example). The way in which the present work relates bicomplexes and conservation laws, however, is different and seems not having been anticipated in the literatur. 
In general, the maps $\d, \delta$ of a bicomplex are not required to be (graded) derivations. In fact, the above iterative construction of $\delta$-closed forms does not make use of the (graded) Leibniz rule. \\
(5) The condition (\ref{del_chi0=0}) can be weakened to $\d \delta \chi^{(0)} =0$. Setting $J^{(0)}=\delta \chi^{(0)}$, this somehow improves the left end of the diagram in Fig.~1.\\
(6) If $H^s_\delta (\O(\A)) \neq \{ 0 \}$, the iterative construction may still work, for some $\chi^{(0)}$, though perhaps only up to some level $m$ where we encounter a $\delta$-closed form $J^{(m)}$ which is not $\delta$-exact.
\hfill                $\diamondsuit$
\vskip.2cm

In this work we will concentrate on the case $s=1$ where $\chi \in \A$. Since $s$-form conservation laws with $s>1$ are of some interest in the theory of differential systems and physical field theories (see \cite{Torr97} and the references given there), we believe that the above generalization has some potential.

\section{Example: a generalization of Plebanski's first heavenly equation}
\label{sec:Plebanski}
\setcounter{equation}{0}
Let $\A$ be the algebra of smooth functions of coordinates $x^\mu$, $\mu = 1, \ldots,2n$, and $y^a$, $a=1,\ldots,2m$, and let $\partial_\mu$ and $\partial_a$ denote the partial derivatives with respect to $x^\mu$ and $y^a$, respectively. We  define
\be
    \delta f = (\partial_\mu f) \, \delta x^\mu
\ee
where the $\delta x^\mu$ are ordinary differentials, which commute with functions, and
\be
    \d f = (M_\mu f) \, \delta x^\mu
\ee
where
\be  
       M_\mu = M_\mu^a \, \partial_a
\ee
with functions $M_\mu^a$. Now $(\delta \d + \d \delta) f=0$ (for all $f \in \A$) means $\delta (M_\mu \, \delta x^\mu) =0$ and thus
\be
     M_\mu^a = \partial_\mu W^a
\ee
with $W^a \in \A$. Furthermore, $\d^2=0$ is satisfied if $[M_\mu , M_\nu]=0$ which leads to
\be
   (\partial_\mu W^a) (\partial_\nu \partial_a W^b) - (\partial_\nu W^a) (\partial_\mu \partial_a W^b) =0 \; .
\ee 
Let us now consider the special case where
\be
     W^a = \omega^{ab} \, \partial_b \O
\ee
with a function $\O$ and constants $\omega^{ab} = - \omega^{ba}$.\footnote{If $\omega^{ab}$ has an inverse $\omega_{ab}$, the latter defines a symplectic 2-form and $W =W^a \partial_a$ is the Hamiltonian vector field associated with the Hamiltonian $\O$.}
Then
\be
   \omega^{ac} \, \omega^{bd} \, \partial_d \{ (\partial_\mu \partial_c \O) (\partial_\nu \partial_a \O) \} =0 \; .
\ee 
If $(\omega^{ab})$ is invertible, this leads to
\be
   \omega^{ab} \, (\partial_\mu \partial_a \O) (\partial_\nu \partial_b \O) = \tilde{\omega}_{\mu \nu}                     \label{mn-heaven}
\ee 
where $\tilde{\omega}_{\mu\nu}$ are arbitrary functions of $x^\mu$, satisfying $\tilde{\omega}_{\mu \nu} = - \tilde{\omega}_{\nu \mu}$. Furthermore, the 2-form 
\be
     \tilde{\omega} = {1 \over 2} \, \tilde{\omega}_{\mu \nu} \, \delta x^\mu \delta x^\nu
\ee
is $\delta$-closed. Let us take $\tilde{\omega}_{\mu\nu}$ to be invertible. Then, by the Darboux theorem, there are local coordinates $x^\mu$ such that
\be
   (\tilde{\omega}_{\mu \nu}) = \left( \begin{array}{cc} 0 & I_n \\ -I_n & 0 \end{array} \right)
\ee
where $I_n$ is the $n \times n$ unit matrix. (\ref{mn-heaven}) generalizes Plebanski's {\em first heavenly equation} \cite{Pleb75} to which it reduces for $m=n=1$:
\be
   \O_{xp} \, \O_{tq} - \O_{xq} \, \O_{tp} = 1
\ee 
where $x^\mu =(t,x)$ and $y^a =(q,p)$. This is a gauge-reduced form of the self-dual gravity equation \cite{Pleb75}.\footnote{A solution $\O$ determines a Riemannian metric with line element $ds^2 = \O_{tp} \, dt \, dp + \O_{tq} \, dt \, dq + \O_{xp} \, dx \, dp + \O_{xq} \, dx \, dq$.}
The above generalization of Plebanski's equation appeared already in \cite{Taka89}.
\vskip.2cm

 For $f,g \in \A$ we introduce the Poisson bracket
\be
    \{ f , g \} = \omega^{ab} \, (\partial_a f) \, (\partial_b g) \; .
\ee
Then
\be
    \d f = \{ f , \delta \Omega \}  \, , \qquad  
    \{ \delta \Omega , \delta \Omega \} = 2 \, \tilde{\omega}    \; .
\ee
\vskip.2cm

The initial condition $\delta \chi^{(0)} =0$ for the iterative construction of $\delta$-closed 1-forms in the preceding section means that $\chi^{(0)} \in \A$ does not depend on $x^\mu$, hence $\chi^{(0)} = \chi^{(0)}(y^a)$. 
\vskip.2cm

 From (\ref{chi_sum_sec3}) and (\ref{chi_eq_sec3}) we get
\be
   J^{(m)} = \delta \chi^{(m)} = \{ \chi^{(m-1)} , \delta \Omega \} \; .
\ee
In particular, this leads to
\be
   \chi^{(1)} = \{ \chi^{(0)} , \Omega \}
\ee
(modulo addition of a function which only depends on $y^a$), so that
\be
   J^{(2)} = \d \chi^{(1)} = \{ \{ \chi^{(0)} , \Omega \} , \delta \Omega \}   \; .
\ee
\vskip.1cm
\noindent
{\em Remark.} Let $\epsilon^{\mu \nu}$ be constant and antisymmetric. We define $\tilde{J}^{(m)\mu} := \epsilon^{\mu\nu} J^{(m)}_\nu$ where $J^{(m)} = J^{(m)}_\mu \, \delta x^\mu$. Now $\delta J^{(m)} =0$ becomes $\partial_\mu \tilde{J}^{(m)\mu} =0$ which is a familiar form of a conservation law. See also \cite{Husa94,Stra95} for related work.

\section{Gauging bi-differential calculi}
\label{sec:gauging bi-diff}
\setcounter{equation}{0}
Let $(\O(\A), \d , \delta)$ be a bi-differential calculus, and $A,B$ two $N \times N$-matrices of 1-forms  (i.e., the entries are elements of $\O^1(\A)$). We introduce two operators (or covariant derivatives)
\be
     D_\d = \d + A \, \qquad  D_\delta = \delta + B
\ee
which act from the left on $N \times N$-matrices with entries in $\O(\A)$. The latter form a graded left $\A$-module ${\cal M} = \bigoplus_{r \geq 0} {\cal M}^r$. Then
\be
      D_\d^2 = 0  \quad &\Longleftrightarrow&  \quad F_\d[A] = \d A + AA =0  
                \label{D2A}   \\
  D_\delta^2 = 0  \quad &\Longleftrightarrow&  \quad F_\delta[B] = \delta B + BB =0  
                \label{D2B}   \\
  D_\d \, D_\delta + D_\delta \, D_\d = 0 \quad &\Longleftrightarrow& \quad  
      \d B + \delta A + BA + AB =0  \; .
                \label{DADB}
\ee
These conditions are sufficient for a generalization of the construction presented in section 2. If they are satisfied, we speak of a {\em gauged bi-differential calculus}.
\vskip.2cm

Suppose there is a (nonvanishing) $\chi^{(0)} \in {\cal M}^{s-1}$ with 
\be
         D_\delta \, \chi^{(0)} =0   \; .
\ee
Then
\be
     J^{(1)} = D_\d \, \chi^{(0)} 
\ee
is $D_\delta$-closed, i.e.
\be
        D_\delta J^{(1)} = - D_\d D_\delta \, \chi^{(0)} = 0   \; .
\ee
If every $D_\delta$-closed element of ${\cal M}^s$ is $D_\delta$-exact, then
\be
       J^{(1)} = D_\delta \, \chi^{(1)}
\ee
with some $\chi^{(1)} \in {\cal M}^{s-1}$. Now let $J^{(m)} \in {\cal M}^s$ satisfy
\be
       D_\delta J^{(m)} = 0 \, , \qquad  J^{(m)} = D_\d \, \chi^{(m-1)}   \; .
\ee 
Then
\be
        J^{(m)} = D_\delta \, \chi^{(m)}     \label{Ddelta_chim}
\ee
with some $\chi^{(m)} \in {\cal M}^{s-1}$ (which is determined only up to addition of some $\beta \in {\cal M}^{s-1}$ with $D_\delta \beta=0$), and
\be
        J^{(m+1)} = D_\d \, \chi^{(m)}  
\ee
is also $D_\delta$-closed:
\be
  D_\delta J^{(m+1)} = - D_\d D_\delta \, \chi^{(m)} = - D_\d J^{(m)} = - D_\d^2 \, \chi^{(m-1)} = 0  \; .
\ee
In this way we obtain an infinite tower (see Fig.~2) of $D_\delta$-closed matrices $J^{(m)}$ of $s$-forms and elements $\chi^{(m)} \in {\cal M}^{s-1}$ which satisfy
\be
      D_\delta \, \chi^{(m+1)} = D_\d \, \chi^{(m)}  \; .   \label{D-chim-eq}
\ee

\diagramstyle[PostScript=dvips]
\begin{diagram}[notextflow]
 &         &\chi^{(0)}&         &       &         &\chi^{(1)}&         &       &         &\chi^{(2)}& & & \\
 &\ldTo^\Dl&          &\rdTo^{D_\d}&       &\ldTo^\Dl&          &\rdTo^{D_\d}&       &\ldTo^\Dl&          &\rdTo^{D_\d}& & \\
0&         &          &         &J^{(1)}&         &          &         &J^{(2)}&         &          & &J^{(3)}&\cdots\\
 &         &          &\ldTo^\Dl&       &\rdTo^{D_\d}&          &\ldTo^\Dl&       &\rdTo^{D_\d}&          &\ldTo^\Dl& &  \\
 &         & 0        &         &       &         & 0        &         &       &         & 0        &&&
\end{diagram}

\vskip.1cm \noindent
\begin{center}
{\bf Fig.~2}  \\
The infinite tower of $D_\delta$-closed (matrices of) $s$-forms $J^{(m)}$.
\end{center}
\vskip.2cm

In terms of 
\be
  \chi = \sum_{m=0}^\infty \lambda^m \, \chi^{(m)}    \label{chi-sum}
\ee
with a parameter $\lambda$, the set of equations (\ref{D-chim-eq}) leads to
\be
    D_\delta \, \chi = \lambda \, D_\d \, \chi  \; .    \label{D-chi-eq}
\ee
Conversely, if the last equation holds for all $\lambda$, we recover (\ref{D-chim-eq}).
\vskip.1cm

Of particular interest is the case $s=1$, as we will demonstrate in section 5. The above procedure works, however, irrespective of this restriction (provided there is a $D_\delta$-closed $s$-form and the cohomology condition is satisfied). It thus opens new possibilities which still have to be explored. The remarks in section 2 apply, with obvious alterations, also to the gauged iteration procedure.
\vskip.2cm

If $B=0$, the  conditions (\ref{D2A})-(\ref{DADB}) become
\be
     F_\d[A] =0 \, , \qquad   \delta A =0  \; .
\ee
There are two obvious ways to further reduce these equations. 
\vskip.2cm
\noindent
(1) We can solve $F_\d[A] =0$ by setting
\be
     A = g^{-1} \, \d g
\ee
with an invertible $N \times N$-matrix $g$ with entries in $\A$. Then the remaining equation reads
\be
     \delta (g^{-1} \, \d g) = 0
\ee
which resembles the field equation of principal chiral models (see also the following section). 
\vskip.2cm
\noindent
(2) We can solve $\delta A = 0$ via
\be
      A = \delta \phi
\ee
with a matrix $\phi$. Then we are left with the equation
\be
      \d (\delta \phi) + (\delta \phi)^2 = 0  \; .
\ee
This generalizes the so-called `pseudodual chiral models' (cf \cite{Curt+Zach94}, see also \cite{Zakh+Mikh78,Napp80}).

\section{Gauged bi-differential calculi and integrable models}
\label{sec:intmod}
\setcounter{equation}{0}
In this section we present a collection of integrable models which arise from gauged bi-differential calculi. As a consequence, they possess an infinite tower of `conserved currents' in the sense of $D_\delta$-closed 1-forms. For some well-known integrable models, like principal chiral models, KP equation and the nonlinear Toda lattice, we show that these reproduce known sets of conserved currents and conserved charges. Moreover, in the last subsection we present a set of equations in $2n$ dimensions which generalize the four-dimensional self-dual Yang-Mills equation and which are integrable in the sense of admitting a gauged bi-differential calculus formulation.

\subsection{Chiral models}
\label{sec:chiral models}
(1) Let $\A = C^\infty(\mathbb{R}^2)$ be the commutative algebra of smooth functions of coordinates $t,x$, and $\delta$ the ordinary exterior derivative acting on the algebra $\O(\A)$ of differential forms on $\mathbb{R}^2$. Then
\be
     \delta f = f_x \, \delta x + f_t \, \delta t  \qquad  \forall f \in \A 
\ee
where $f_x$ and $f_t$ denote the partial derivatives of $f$ with respect to $x$ and  $t$, respectively. As a consequence of the Poincar{\'e} Lemma, every $\delta$-closed 1-form is $\delta$-exact.  
An extension of this differential calculus to a bi-differential calculus is obtained by defining another differential map $\d$ via
\be   
     \d f = f_t \, \delta x + f_x \, \delta t   \, , \qquad
     \d (f \, \delta x + h \, \delta t) = (\d f) \, \delta x + (\d h) \, \delta t  \; .
\ee
Indeed, we have
\be
  \d \delta f &=& (f_{xx} - f_{tt}) \, \delta t \, \delta x = - \delta \d f  \\
  \d^2 f  &=& (f_{tx}-f_{xt}) \, \delta t \, \delta x = 0
\ee
and $\d$ also satisfies the graded Leibniz rule (\ref{Leibniz}). 
Now $F_\d[A] =0$ is solved by 
\be
    A = g^{-1} \, \d g = g^{-1} g_t \, \delta x + g^{-1} g_x \, \delta t  
\ee
with an invertible $N \times N$-matrix $g$ with entries in $\A$. With $B=0$, the remaining condition (\ref{DADB}) for a gauged bi-differential calculus is $\delta A =0$ which turns out to be equivalent to the principal chiral model equation
\be
    (g^{-1} g_t)_t = (g^{-1} g_x)_x  \; . 
\ee
It has the form of a conservation law. More generally, $\delta J =0$ for a 1-form $J = J_0 \, \delta t + J_1 \, \delta x$ is equivalent to the conservation law $J_{1,t} = J_{0,x}$. Hence $Q = \int_{\mbox{\small $t=$ const.}} J$ is conserved (if $J_0$ vanishes sufficiently fast at spatial infinity). From (\ref{Ddelta_chim}) we get
\be
   J = \sum_{m=1}^\infty \lambda^m \, J^{(m)} = \lambda \, D_\d \chi
     = \lambda \, \left[ ( \chi_t + g^{-1} g_t \, \chi ) \, \delta x 
         + ( \chi_x + g^{-1} g_x \, \chi ) \, \delta t \right]
\ee
and $\delta J=0$ leads to
\be
   ( \chi_t + g^{-1} g_t \, \chi )_t = ( \chi_x + g^{-1} g_x \, \chi )_x  \; . 
\ee
\vskip.1cm

(\ref{D-chim-eq}) takes the form
\be
    \chi_t = \lambda \, (\chi_x + g^{-1} g_x \, \chi) \, , \qquad
    \chi_x = \lambda \, (\chi_t + g^{-1} g_t \, \chi)   \; .
\ee
Inserting (\ref{chi-sum}) with $\chi^{(0)}=I$, the $N \times N$ unit matrix\footnote{This satisfies $\delta \chi^{(0)}=0$. The most general solution of $\delta \chi^{(0)}=0$ is the $N \times N$ matrix where the entries are arbitrary constants. Instead of the $Q$ obtained from the initial data $\chi^{(0)}=I$, we then simply get $Q$ multiplied from the right by this general $N \times N$ matrix.}, in the last equation, we obtain the conserved charges
\be
   Q^{(1)} &=& \int_{\mbox{\small $t=$ const.}} g^{-1} g_t \, \delta x  \\
   Q^{(2)} &=& \int_{\mbox{\small $t=$ const.}} 
         ( \chi^{(1)}_t + g^{-1} g_t \, \chi^{(1)} ) \, \delta x
                  \nonumber \\
       &=& \int_{\mbox{\small $t=$ const.}} 
         \left( g^{-1} g_x + g^{-1} g_t \, \int^x g^{-1} g_t \, \delta x' \right) \, \delta x        
\ee
and so forth. In this way one recovers the infinite tower of nonlocal conserved charges for two-dimensional principal chiral models \cite{BIZZ79}.
\vskip.2cm
\noindent
(2) Let $\A = C^\infty(\mathbb{R}^3)$ with coordinates $t,x,y$. Regarding $x$ as a parameter, the ordinary calculus of differential forms on the algebra of smooth functions of $t$ and $y$ induces a differential calculus $(\O(\A),\delta )$ such that
\be
      \delta f = f_t \, \delta t + f_y \, \delta y   \; .
\ee
Now
\be
  \d f = f_x \, \delta t + f_t \, \delta y  \, , \qquad
  \d (f \, \delta t + h \, \delta y) = (\d f) \, \delta t + (\d h) \, \delta y
\ee
defines a map $\d$ satisfying the graded Leibniz rule, $\d^2 =0$ and $\d \delta = - \delta \d$. 
With $A=g^{-1} \d g$ we have $F_\d[A] = 0$, and (with $B=0$) the condition $\delta A=0$ becomes
\be
      (g^{-1} g_t)_t = (g^{-1} g_x)_y  \; . 
\ee
 From $\delta (\d \chi + A \chi) =0$ (which is an integrability condition of (\ref{D-chi-eq}) ), one obtains the conservation law
\be
    (\chi_t + g^{-1} g_t \, \chi)_t = (\chi_y)_x + (g^{-1} g_x \, \chi)_y
\ee
which leads to the conserved quantity
\be
  Q = \int_{\mbox{\small $t=$ const.}} (\chi_t + g^{-1} g_t \, \chi) \, \delta x \, \delta y 
\ee
(assuming that $g^{-1} g_x$ and $\chi_y$ vanish sufficiently fast at spatial infinity). Furthermore, (\ref{D-chi-eq}) takes the form $\delta \chi = \lambda \, (\d + A) \chi$ which leads to
\be
    \chi_t = \lambda \, (\chi_x + g^{-1} g_x \, \chi) \, , \qquad
    \chi_y = \lambda \, (\chi_t + g^{-1} g_t \, \chi)   \; .
\ee
Using (\ref{chi-sum}) with $\chi^{(0)}=I$, the $N \times N$ unit matrix, we obtain the conserved charges
\be
  Q^{(1)} &=& \int_{\mbox{\small $t=$ const.}} g^{-1} g_t \, \delta x \, \delta y  \\
  Q^{(2)} &=& \int_{\mbox{\small $t=$ const.}} \left( g^{-1} g_x 
   + g^{-1} g_t \, \int^y g^{-1} g_t \, \delta y' \right) \, \delta x \, \delta y
\ee
and so forth.

\subsection{Toda models}
\label{sec:Toda models}
(1) Let $\A$ be the algebra of functions of $t,k,S,S^{-1}$ which are smooth in $t$ and formal power series in the shift operator
\be
     S(f)_k = f_{k+1}
\ee 
and its inverse $S^{-1}$. $k$ has values in $\mathbb{Z}$ and we introduced the notation $f_k(t,S,S^{-1}) = f(t,k,S,S^{-1})$. Because of the relations $S \, f_k = f_{k+1} \, S$ and $S^{-1} \, f_k = f_{k-1} \, S^{-1}$, the algebra $\A$ is noncommutative. We define a bi-differential calculus over $\A$ via
\be
    \delta f = \dot{f} \, \delta t + [ S , f ] \, \xi    \, , \qquad
    \d f = [ S^{-1} , f ] \, \delta t - \dot{f} \, \xi  
\ee
where $(\delta t)^2 = 0 = \xi^2$, $\xi \, \delta t + \delta t \, \xi =0$ and $\dot{f} = \partial f/\partial t$. $\delta t$ and $\xi$ commute with all elements of $\A$. The action of $\delta$ extends to 1-forms via
\be
  \delta (f \, \delta t + h \, \xi) = (\delta f) \, \delta t + (\delta h) \, \xi
\ee
and correspondingly for $\d$. Indeed,
\be
  \d^2 f &=& -(\d \dot{f}) \, \xi + \d [ S^{-1} , f ] \, \delta t 
  = -[ S^{-1} , \dot{f} ] \, \delta t \, \xi - [ S^{-1} , f ]\dot{ } 
      \, \xi \, \delta t = 0  \\
    \d \delta f &=& (\d \dot{f}) \, \delta t + \d [ S , f ] \, \xi 
  = -\ddot{f} \, \xi \, \delta t + [ S^{-1} , [ S , f ] ] \, \delta t \, \xi \nonumber \\
    &=& \ddot{f} \, \delta t \, \xi - [ S , [ S^{-1} , f ] ] \, \xi \, \delta t
  = - \delta \d f 
\ee
and similar calculations demonstrate that the rules of bi-differential calculus are satisfied.
Let
\be
   A = e^{-q_k} \, \d e^{q_k} = (e^{q_{k-1}-q_k} - 1) \, S^{-1} \, \delta t - \dot{q}_k \, \xi
\ee
with a function $q_k(t) = q(t,k)$ and $\dot{q}_k = \partial q_k/\partial t$. Then $F_\d[A] = 0$ and, using
\be
    [ S , (e^{q_{k-1}-q_k} - 1) \, S^{-1} ] = [ S , e^{q_{k-1}-q_k} ] \, S^{-1} 
    = ( e^{q_k-q_{k+1}} - e^{q_{k-1}-q_k} ) \, S \, S^{-1} \, ,
\ee
we recover from $\delta A = 0$ (thus setting $B=0$) the nonlinear Toda lattice equation \cite{Toda81}
\be
    \ddot{q}_k = e^{q_{k-1}-q_k} - e^{q_k-q_{k+1}}   \; .
\ee 
\vskip.1cm

(\ref{D-chi-eq}) is equivalent to the system
\be
  \dot{\chi}_k &=& \lambda \, \left( e^{q_{k-1}-q_k}  \,
                  \chi_{k-1} - \chi_k  \right) \, S^{-1}   \\
  \chi_{k+1}-\chi_k &=& - \lambda \, ( \dot{\chi}_k + \dot{q}_k \chi_k ) \, S^{-1}
\ee
which leads to
\be
   \chi_{k+1}-\chi_k &=& - \lambda \, \dot{q}_k \chi_k \, S^{-1}
  + \lambda^2 \, (\chi_k - e^{q_{k-1}-q_k} \, \chi_{k-1} ) \, S^{-2} \; .
\ee
Inserting
\be
   \chi_k = \sum_{m=0}^\infty \lambda^m \, \tilde{\chi}^{(m)}_k \, S^{-m}
\ee
with $\tilde{\chi}^{(0)}=1$ in the last equation leads to 
\be
    \tilde{\chi}^{(1)}_{k+1} - \tilde{\chi}^{(1)}_k = - \dot{q}_k
\ee
and
\be
   \tilde{\chi}^{(m)}_{k+1} - \tilde{\chi}^{(m)}_k 
 = - \dot{q}_k \, \tilde{\chi}^{(m-1)}_k + \tilde{\chi}^{(m-2)}_k
   - e^{q_{k-1}-q_k} \, \tilde{\chi}^{(m-2)}_{k-1}    \label{chitilde_diff}
\ee
for $m>1$. Hence
\be
    \tilde{\chi}^{(1)}_k = - \sum_{j=-\infty}^{k-1} \dot{q}_j
\ee
(provided that the infinite sum on the r.h.s. exists) and
\be
   \tilde{\chi}^{(m)}_k 
 = \sum_{j=-\infty}^{k-1} \left(- \dot{q}_j \, \tilde{\chi}^{(m-1)}_j 
   + \tilde{\chi}^{(m-2)}_j - e^{q_{j-1}-q_j} \, \tilde{\chi}^{(m-2)}_{j-1} \right)
                   \label{chi^m-Toda}
\ee
for $m>1$. In particular,
\be
   \tilde{\chi}^{(2)}_k 
 = \sum_{j=-\infty}^{k-1} \left( - \dot{q}_j \, \tilde{\chi}^{(1)}_j 
   + 1 - e^{q_{j-1}-q_j} \right) 
 = \sum_{j=-\infty}^{k-1} \sum_{l=-\infty}^{j-1} \dot{q}_j \, \dot{q}_l 
   + \sum_{j=-\infty}^{k-1} \left( 1 - e^{q_{j-1}-q_j} \right)  \; .
\ee
\vskip.1cm

 For a 1-form $J = J_0 \, \delta t + J_1 \, S \, \xi$ the condition $\delta J =0$
reads $\dot{J}_1 = S(J_0)-J_0 = \partial_+ J_0$ where the r.h.s. is the discrete forward derivative of $J_0$.\footnote{Setting $\chi = \int J_0 \, dt$ (ordinary integration with respect to $t$), one easily verifies that $\delta J=0$ implies $J=\delta \chi$, so that $\delta$-closed 1-forms are $\delta$-exact.} The latter equation is a conservation law. Indeed,
for
\be
   Q = \int_{\mbox{\small $t=$ const.}} J = \sum_{k=-\infty}^\infty J_{1k} 
\ee
where the last equality defines the integral, we have
\be
  {d \over dt} Q = \int_{\mbox{\small $t=$ const.}} (\partial_+ J_0) \, S \, \xi 
                 = 0  
\ee
if $ J_{0k} $ vanishes sufficiently fast for $ k \to \pm \infty $.
Using $J^{(m)} = \tilde{J}^{(m)} \, S^{-m}$ and (\ref{Ddelta_chim}), we find
\be
   \tilde{Q}^{(m)} &=& \int_{\mbox{\small $t=$ const.}} \tilde{J}^{(m)} 
  = \int_{\mbox{\small $t=$ const.}} \delta \tilde{\chi}^{(m)}   \nonumber \\
 &=& \sum_{k=-\infty}^\infty \left( - \dot{q}_k \, \tilde{\chi}^{(m-1)}_k 
    + \tilde{\chi}^{(m-2)}_k - e^{q_{k-1}-q_k} \, \tilde{\chi}^{(m-2)}_{k-1} \right)
\ee
which, together with (\ref{chi^m-Toda}), allows the recursive calculation of the conserved 
charges $\tilde{Q}^{(m)}$.\footnote{By using (\ref{chitilde_diff}) we also have 
$\tilde{Q}^{(m)} = \tilde{\chi}^{(m)}_\infty - \tilde{\chi}^{(m)}_{-\infty} $.}
In particular, we get
\be
    - \tilde{Q}^{(1)} = \sum_{k=-\infty}^\infty \dot{q}_k 
\ee 
and 
\be
    {1 \over 2} \, (\tilde{Q}^{(1)})^2 - \tilde{Q}^{(2)} 
 = {1 \over 2} \, \sum_{k=-\infty}^\infty \dot{q}_k^2 
   + \sum_{k=-\infty}^\infty ( e^{q_{k-1}-q_k} -1 ) 
\ee
which are the total momentum and total energy, respectively. Proceeding further with the iteration, one recovers the higher conserved charges of the Toda lattice as given, for example, in \cite{Heno74}. For instance, introducing $X_k = e^{q_{k-1}-q_k}$ we find
\be
   \tilde{Q}^{(3)} &=& - \sum_{k=-\infty}^\infty \dot{q}_{k-1} \, X_k 
  -  \sum_{k=-\infty}^\infty \sum_{j=-\infty}^{k-1} \sum_{l=-\infty}^{j-1} 
     \dot{q}_k \, \dot{q}_j \, \dot{q}_l  \nonumber \\
 & &  + \sum_{k=-\infty}^\infty \sum_{j=-\infty}^{k-1} \left( \dot{q}_j \, (X_k - 1) 
      +  \dot{q}_k \, (X_j - 1) \right)  
\ee
and after some resummations 
%using the identities
%\be
%\sum_{k=-\infty}^\infty \sum_{j=-\infty}^{k-1} \sum_{l=-\infty}^{j-1}
%     \dot{q}_k \, \dot{q}_j \, \dot{q}_l 
%   = {1 \over 6} \, [-\tilde{Q}^{(1)}]^3
%     - {1\over2} \, [-\tilde{Q}^{(1)}] \, \sum_{k=-\infty}^\infty \dot{q}_k^2 
%     + {1\over3} \, \sum_{k=-\infty}^\infty \dot{q}_k^3    
%\ee
%and
%\be
%  \sum_{k=-\infty}^\infty \sum_{j=-\infty}^{k-1} \left( 
%       \dot{q}_j \, (X_k-1) + \dot{q}_k \, (X_j-1) \right) 
%   = [-\tilde{Q}^{(1)}] \sum_{k=-\infty}^\infty (X_k-1)
%      - \sum_{k=-\infty}^\infty \dot{q}_k \, (X_k-1)  
%\ee
we obtain the formula
\be
  - \tilde{Q}^{(3)} + \tilde{Q}^{(1)} \, \tilde{Q}^{(2)} - \tilde{Q}^{(1)} - {1\over3} \, [\tilde{Q}^{(1)}]^3 
  = \sum_{k=-\infty}^\infty \left( {1\over3} \, \dot{q}_k^3 + \dot{q}_k \, (X_k + X_{k+1}) \right)  \; .
\ee
 
\vskip.2cm
\noindent
{\em Remark:} Since $\delta k = S \, \xi$, we have $[\delta k , f] = ( S(f)-f ) \, \delta k$ and $\delta f = \dot{f} \, \delta t + ( S(f)-f ) \, \delta k$ for functions $f(t,k)$. Since $\delta k$ does not in general commute with functions, the last two equations define a noncommutative differential calculus over the commutative algebra of functions on $\mathbb{R} \times \mathbb{Z}$ \cite{DMHS93}. There is an integral naturally associated with this calculus. It satisfies
\bez
      \int_{\mathbb{Z}} f(t,k) \, \delta k = \sum_{k=-\infty}^\infty f(t,k) \; .
\eez
We refer to \cite{DMH92} for details. See also \cite{DMH96} for a different derivation of the conserved charges for the Toda lattice in this framework.
\hfill                $\diamondsuit$
\vskip.2cm
\noindent
(2) A generalization of the previous example is obtained as follows. Let $\A$ be the algebra of functions of $t,x,k,S,S^{-1}$ which are smooth in $t$ and $x$, and polynomial in the shift operators $S, S^{-1}$. Again, $k$ has values in $\mathbb{Z}$. A bi-differential calculus over $\A$ is then obtained via
\be
    \delta f = \dot{f} \, \delta t + [ S , f ] \, \xi    \, , \qquad
    \d f = [ S^{-1} , f ] \, \delta t + f' \, \xi
\ee
where $f'=\partial f/\partial x$. With
\be
   A = e^{-q_k} \, \d e^{q_k} = (e^{q_{k-1}-q_k} - 1) \, S^{-1} \, \delta t + q'_k \, \xi
\ee
we have $F_\d[A] = 0$ and $\delta A = 0$ becomes the Toda field equation
\be
    \dot{q}'_k = e^{q_k-q_{k+1}} - e^{q_{k-1}-q_k}  \; .
\ee 
Alternatively, we can solve $\delta A=0$ by $A=\delta (u \, S^{-1})$ with a function $u(t,x,k)$. Then $F_\d[A] = 0$ reads
\be
   \dot{u}' + (1+\dot{u}) \, \Delta u = 0
\ee
where $\Delta u = S(u)+S^{-1}(u) - 2 u$. The latter equation has been studied in \cite{HIK88}.
\vskip.2cm
\noindent
(3) Let $\A$ be as in the previous example and consider the bi-differential calculus determined by
\be
    \delta f = [ S^{-1} , f ] \, \tau - f' \, \xi    \, , \qquad
    \d f = \dot{f} \, \tau + [ S , f ] \, \xi  \; .   
\ee
With
\be
       A = X \, \tau + (Y-I) S \, \xi
\ee
where $X,Y$ are matrices with entries in $\A$ and $I$ is the unit matrix, $\delta A = 0$ leads to
\be
       {X_k}' = Y_k - Y_{k-1}
\ee 
and $F_\d[A]=0$ becomes
\be
      \dot{Y}_k = Y_k \, X_{k+1} - X_k \, Y_k   \; .
\ee
 For $x=t$, the (transpose of the) last two equations are those of the non-Abelian Toda lattice explored in \cite{Gekh98}, for example.

\subsection{The KP equation}
\label{sec:KP equation}
Let $\A_0 = C^\infty(\mathbb{R}^3)$ be the algebra of smooth functions of coordinates $t, x, y$, and $\A$ the algebra of formal power series in the partial derivative $\partial_x = \partial/\partial x$ with coefficients in $\A_0$. We define a bi-differential calculus over $\A$ via
\be
  \d f &=& [ \partial_t - \partial_x^3 , f ] \, \tau 
          + [ {1 \over 2} \partial_y - {1 \over 2} \partial_x^2 , f ] \, \xi  \nonumber \\
       &=& ( f_t - f_{xxx} - 3 \, f_{xx} \, \partial_x 
          - 3 \, f_x \, \partial_x^2 ) \, \tau 
          + {1 \over 2} \, ( f_y - f_{xx} - 2 \, f_x \, \partial_x ) \, \xi  \\
 \delta f &=& [ {3 \over 2} \partial_y + {3 \over 2} \partial_x^2 , f ] \, \tau 
             + [ \partial_x , f ] \, \xi 
           = {3 \over 2} \, ( f_y + f_{xx} + 2 \, f_x \, \partial_x ) \, \tau + f_x \, \xi   \; .
\ee
 For a gauge potential $A \in \O^1(\A)$ we solve the equation $\delta A=0$ by 
\be
   A = \delta v = {3 \over 2} ( v_y + v_{xx} + 2 \, v_x \, \partial_x ) \, \tau + v_x \, \xi
\ee
with $v \in \A_0$. Then $F_\d[A] =0$ takes the form
\be
    v_{xt} - {1 \over 4} \, v_{xxxx} + 3 \, v_x \, v_{xx} - {3 \over 4} \, v_{yy} = 0 \; .
                   \label{KPa}
\ee
Differentiation with respect to $x$ and substitution $u = - v_x$ leads to the KP equation
\be
    (u_{t} - {1 \over 4} \, u_{xxx} - 3 \, u u_x)_x - {3 \over 4} \, u_{yy} = 0 
\ee
in the form considered, for example, in \cite{MSS90}.
\vskip.2cm

Let us now turn to the conservation laws. First we note that the integrability condition $\delta D_\d \, \chi = 0$ of (\ref{D-chi-eq}) for $\chi \in {\cal M}$ can be written in the form of a conservation law,
\be
   (\chi_x)_t = {3 \over 4} \, ( \chi_y + 2 \, v_x \, \chi )_y + ( {1 \over 4} \, \chi_{xxx}
   - {3 \over 2} \, v_y \, \chi - {3 \over 2} \, v_x \, \chi_x  )_x  \; .
\ee
Note that terms proportional to $\partial_x$ cancel each other in the evaluation of $\delta D_\d \, \chi$ . 
Moreover, in the case under consideration (\ref{D-chi-eq}) consists of the two equations
\be
    \chi_x = \lambda \, ( {1 \over 2} \, \chi_y - {1 \over 2} \, \chi_{xx} + v_x \, \chi 
             - \chi_x \, \partial_x )
\ee
and
\be
    \chi_y + \chi_{xx} + 2 \, \chi_x \, \partial_x 
  &=& \lambda \, [ {2 \over 3} \, ( \chi_t - \chi_{xxx} - 3 \, \chi_{xx} \, \partial_x 
     - 3 \, \chi_x \, \partial_x^2 )   \nonumber \\
  & &  + v_y \, \chi + v_{xx} \, \chi + 2 \, v_x \, \chi_x 
       + 2 \, v_x \, \chi \, \partial_x ]   \; .
\ee
Inserting 
\be
      \chi = \sum_{n=0}^\infty \chi_n \, \partial_x^n 
\ee
we get
\be
  \chi_{0,x} &=& {\lambda \over 2} \, ( \chi_{0,y} - \chi_{0,xx} + 2 \, v_x \, \chi_0 ) \\
  \chi_{0,y} + \chi_{0,xx} 
 &=& \lambda \, [ {2 \over 3} \, ( \chi_{0,t} - \chi_{0,xxx} )
     + v_y \, \chi_0 + v_{xx} \, \chi_0 + 2 \, v_x \, \chi_{0,x} ] \; .  \label{KP-chi0-2}
\ee
The transformation
\be
    \chi_0 = e^{\lambda \, \varphi} \, , \qquad  
    \varphi = \sum_{m=0}^\infty \lambda^m \, \varphi^{(m)} 
\ee
(which sets $\chi_0^{(0)}=1$) in the first of these equations yields
\be 
  \varphi_x = {\lambda \over 2} \, ( \varphi_y - \varphi_{xx} ) 
              - {\lambda^2 \over 2} \, (\varphi_x)^2 - u
\ee
which in turn leads to
\be
  \varphi^{(0)}_x &=& - u    \label{varphi0} \\  
  \varphi^{(1)}_x &=& - {1 \over 2} \, \partial_x^{-1} u_y + {1 \over 2} \, u_x     \\
  \varphi^{(2)}_x &=& - {1 \over 2} \, u^2 - {1 \over 4} \, u_{xx} 
    + {1 \over 2} \, u_y - {1 \over 4} \, \partial_x^{-2} u_{yy}  
              \label{varphi2}  \\
  \varphi^{(3)}_x &=& - {1 \over 2} \, u \, \partial_x^{-1} u_y 
    - {1 \over 4} \, \partial_x^{-1} (u^2)_y
    + {1 \over 2} \, (u^2)_x - {1 \over 8} \, \partial_x^{-3} u_{yyy} \nonumber \\
  & & + {3 \over 8} \, \partial_x^{-1} u_{yy} - {3 \over 8} \, u_{xy} 
      + {1 \over 8} \, u_{xxx}   \label{varphi3}
\ee
and so forth, where $\partial_x^{-1}$ formally indicates an integration with respect to $x$. 
These are conserved densities of the KP equation (cf \cite{MSS90}\footnote{(\ref{varphi0})-(\ref{varphi3}) correspond to equations (4.15a-d) in \cite{MSS90}. (4.15c) and (4.15d) contain misprints, however. The correct expressions are obtained from the appendix in \cite{MSS90} together with (4.14a-c).}). 
Indeed, in terms of $\varphi$, equation (\ref{KP-chi0-2}) reads
\be
   \varphi_t &=& [ {3 \over 2 \lambda} ( \varphi - v) ]_y + [ \varphi_{xx} 
     + {3 \over 2 \lambda} \, (\varphi - v)_x + {3 \over 2} \, \lambda \, (\varphi_x)^2 ]_x
       \nonumber \\
  & & - 3 \, v_x \, \varphi_x + {3 \over 2} \, (\varphi_x)^2 
      + \lambda^2 \, (\varphi_x)^3 \;.
\ee
Differentiation with respect to $x$ now leads to a conservation law for $\varphi_x$.
\vskip.1cm

We still have to check that $\delta$-closed 1-forms are $\delta$-exact. $\delta J=0$ with $J = J_0 \, \tau + J_1 \, \xi$ means $J_{0,x} = (3/2) (J_{1,y} + J_{1,xx} + 2 \, J_{1,x} \, \partial_x )$. Then  $J = \delta (\partial_x^{-1} J_1)$.

\subsection{Sine-Gordon and Liouville equation}
\label{sec:sine-Gordon}
Let $\A = C^\infty(\mathbb{R}^2)$ be the commutative algebra of smooth functions of coordinates $u$ and $v$, and $\delta$ the ordinary exterior derivative acting on the algebra $\O(\A)$ of differential forms on $\mathbb{R}^2$. Then
\be
     \delta f = f_u \, \delta u + f_v \, \delta v  \qquad  \forall f \in \A  
\ee
where $f_u$ and $f_v$ denote the partial derivatives of $f$ with respect to $u$ and  $v$, respectively. Another differential map $\d$ is then given by
\be   
     \d f = - f_u \, \delta u + f_v \, \delta v   \, , \qquad
     \d (f \, \delta u + h \, \delta v) = (\d f) \, \delta u + (\d h) \, \delta v
\ee
and $(\O(\A), \d , \delta)$ becomes a bi-differential calculus. It is convenient to introduce the 1-forms
\be
    \alpha = \lambda \, \delta u + \lambda^{-1} \, \delta v \, , \qquad
    \beta = - \lambda \, \delta u + \lambda^{-1} \, \delta v
            \label{alpha-beta}
\ee
with a parameter $\lambda$. They satisfy\footnote{Actually, (\ref{alpha-beta}) is the most general solution of these  equations.}
\be
    (\delta f) \, \alpha = - (\d f) \, \beta \, , \quad 
    (\delta f) \, \beta = - (\d f) \, \alpha \, , \quad
    \alpha \, \beta = 2 \, \delta u \, \delta v  \; .
\ee
Let $X_a$, $a=1,2,3$, be a representation of $sl(2)$:
\be
    [X_1 , X_2 ] = X_3 \, ,  \quad 
    [X_1 , X_3 ] = X_2 \, , \quad 
    [X_2 , X_3 ] = X_1   \; .
\ee
\vskip.2cm
\noindent
(1) Now we choose $A = A^a \, X_a$ with
\be
    A^1 = (\cos {\varphi \over 2}) \, \beta \, , \quad 
    A^2 = {1 \over 2} \, \delta \varphi \, , \quad
    A^3 = - (\sin {\varphi \over 2}) \, \alpha \; .
\ee
Then $F_\d[A] =0$ is equivalent to the sine-Gordon equation
\be
     \varphi_{uv} = \sin \varphi  \; .  \label{sineG}
\ee 
Similarly, let $B = B^a \, X_a$ where
\be
    B^1 = - (\cos {\varphi \over 2}) \, \alpha \, , \quad 
    B^2 = {1 \over 2} \, \d \varphi \, , \quad
    B^3 = (\sin {\varphi \over 2}) \, \beta  \; .
\ee
Again, $F_\delta[B] =0$ is equivalent to the above sine-Gordon equation. Moreover, (\ref{DADB}) is satisfied.
Let us now consider the following nonlinear realization of $sl(2)$,
\be
 \tilde{X}_1 = -2 \, \sin {\psi \over 2} \, {\partial \over \partial \psi} \, , \quad
 \tilde{X}_2 = -2 \, {\partial \over \partial \psi} \, , \quad
 \tilde{X}_3 = -2 \, \cos {\psi \over 2} \, {\partial \over \partial \psi} \; .
\ee
To start the iteration procedure of section 4, we need some $\psi=\chi^{(0)}$ with $D_\delta \psi =0$.
With $\tilde{B} = - B^a \, \tilde{X}_a$, this condition becomes $\delta \psi + \tilde{B} \, \psi =0$, respectively
\be
    \delta \psi + \d \varphi &=& 
   2 \, \sin {\psi \over 2} \, \cos {\varphi \over 2} \, \alpha 
 - 2 \, \cos {\psi \over 2} \, \sin {\varphi \over 2} \, \beta
   \nonumber \\
    &=& 2 \lambda \, \sin {\psi + \varphi \over 2} \, \delta u 
    + {2 \over \lambda} \, \sin {\psi - \varphi \over 2} \, \delta v   \; .   \label{preBT}
\ee
Acting with $\delta$ on this equation leads to
\be
    \delta \d \varphi = ( \delta \psi + \d \varphi ) \, ( \cos {\psi \over 2} \, \cos {\varphi \over 2} \, \alpha + \sin {\psi \over 2} \, \sin {\varphi \over 2} \, \beta )
  = 2 \, \sin \varphi \, \delta u \, \delta v 
\ee
which is the sine-Gordon equation (\ref{sineG}) for $\varphi$. In the same way, acting with $\d$ on (\ref{preBT}) leads to the sine-Gordon equation for $\psi$, i.e., $\psi_{uv} = \sin \psi$. Decomposed in the basis $\delta u, \delta v$, (\ref{preBT}) becomes
\be
    (\psi - \varphi)_u = 2 \lambda \, \sin {\psi+\varphi \over 2}  \, , \qquad
    (\psi + \varphi)_v = {2 \over \lambda} \, \sin {\psi-\varphi \over 2}
\ee
which is a well-known B\"acklund transformation for the sine-Gordon equation (see \cite{Draz+John89}, for example).\footnote{The sine-Gordon equation also appeared in treatments of the SU(2) and O(3) chiral models \cite{Zakh+Mikh78,Pohl76}. Our approach above is not related to the discussion of the chiral model in subsection 5.1 in such a way.} 
\vskip.2cm
\noindent
(2) Now we set
\be
    A^1 = \delta \varphi \, , \quad 
    A^2 = e^\varphi \, \alpha  \, , \quad
    A^3 = e^\varphi \, \beta   \; .
\ee
Then $F_\d[A]=0$ with $A = A^a \, X_a$ is equivalent to the Liouville equation
\be
    \varphi_{uv} = e^{2 \varphi}  \; .   \label{Liouville}
\ee
Also $F_\delta[B]=0$ with $B = B^a \, X_a$ and
\be
    B^1 = \d \varphi \, , \quad 
    B^2 = e^\varphi \, \beta  \, , \quad
    B^3 = e^\varphi \, \alpha 
\ee
is equivalent to (\ref{Liouville}). Let us now consider the following nonlinear realization of $sl(2)$,
\be
   \tilde{X}_1 = {\partial \over \partial \psi} \, , \quad
   \tilde{X}_2 = \cosh \psi \, {\partial \over \partial \psi} \, , \quad
   \tilde{X}_3 = \sinh \psi \, {\partial \over \partial \psi} \; .
\ee
With $\tilde{B} = - B^a \, \tilde{X}_a$, the equation $\delta \psi + \tilde{B} \psi =0$ becomes
\be
    \delta \psi - \d \varphi 
  = e^\varphi \, \left( \sinh \psi \, \alpha + \cosh \psi \, \beta \right) \; .
    \label{preBL-L}
\ee 
Acting with $\delta$ on this equation yields
\be
    \delta \d \varphi 
  = - e^\varphi \, ( \delta \psi - \d \varphi ) \, 
    \left( \cosh \psi \, \alpha + \sinh \psi \, \beta \right) 
  = 2 \, e^{2 \varphi} \, \delta u \, \delta v 
\ee 
which reproduces the Liouville equation (\ref{Liouville}). Acting with $\d$ on (\ref{preBL-L}) leads to $\d \delta \psi =0$ and thus $\psi_{uv}=0$. Decomposition of (\ref{preBL-L}) yields
\be
    (\psi + \varphi)_u = - \lambda \, e^{\varphi-\psi} \, , \qquad
    (\psi - \varphi)_v = \lambda^{-1} e^{\varphi+\psi}
\ee
which is a well-known B\"acklund transformation for the Liouville equation (cf \cite{Draz+John89}, for example).
\vskip.2cm

There is a way to construct an infinite set of conserved currents from a given conservation law (like energy conservation) with the help of the B\"acklund transformation (see \cite{SCM73}, for example). So far we have not been able to establish a more direct realization of such conserved quantities within our framework.

\subsection{Self-dual Yang-Mills equations in $2n$ dimensions}
\label{sec:sdYM equations}
Let $(\O(\A),\d)$ and $(\bar{\O}(\A),\bar{\d})$ be two differential calculi over $\A$ such that there is a bijection $\kappa \, : \, \O(\A) \rightarrow \bar{\O}(\A)$ with
\be
    \kappa (w w') = \kappa(w) \, \kappa(w') \qquad \forall w,w' \in \O(\A)
\ee
and $\kappa$ restricted to $\A$ is the identity. Then $\delta = \kappa^{-1} \circ \bar{\d} \circ \kappa$ extends $(\O(\A),\d)$ to a bi-differential calculus, provided that $\d \delta + \delta \d =0$ holds.
\vskip.2cm

Now we choose $\A$ as the algebra of smooth functions of coordinates $x^\mu$, $\mu = 1, \ldots, n$, and $x^{\bar \mu}$, $\bar{\mu}=1, \ldots, n$. Let $(\hat{\O}(\A),\hat{\d})$ denote the ordinary differential calculus over $\A$.
We introduce an invertible $\A$-linear map $\star \, : \, \hat{\O}^2(\A) \rightarrow \hat{\O}^2(\A)$ via
\be
  \star \, ( \hat{\d} x^\mu \, \hat{\d} x^\nu ) 
      &=& - \hat{\d} x^\mu \, \hat{\d} x^\nu  \\
  \star \, ( \hat{\d} x^{\bar{\mu}} \, \hat{\d} x^{\bar{\nu}} ) 
      &=& - \hat{\d} x^{\bar{\mu}} \, \hat{\d} x^{\bar{\nu}}  \\
  \star \, ( \hat{\d} x^\mu \, \hat{\d} x^{\bar{\nu}} ) 
      &=& \kappa^\mu{}_{\bar{\sigma}} \,  \kappa^{\bar{\nu}}{}_\rho 
          \, \hat{\d} x^\rho \, \hat{\d} x^{\bar{\sigma}} 
\ee
where $(\kappa^\mu{}_{\bar \nu})$ is an invertible matrix of constants with inverse $(\kappa^{\bar \nu}{}_\mu)$.
Let 
\be
    \hat{A} = A_\mu \, \hat{\d} x^\mu + B_{\bar{\mu}} \, \hat{\d} x^{\bar{\mu}}
\ee
be a gauge potential ($N \times N$-matrix of 1-forms) with curvature
\be
  F_{\hat{\d}}[\hat{A}] = \hat{\d} \hat{A} + \hat{A} \, \hat{A} 
  = {1 \over 2} \, F_{\mu\nu} \, \hat{\d} x^\mu \, \hat{\d} x^\nu
    + {1 \over 2} \, F_{\bar{\mu}\bar{\nu}} \, \hat{\d} x^{\bar{\mu}} \, \hat{\d} x^{\bar{\nu}}
    + F_{\mu \bar{\nu}} \, \hat{\d} x^\mu \, \hat{\d} x^{\bar{\nu}}  \, ,
\ee
on which we impose the following generalized self-dual Yang-Mills equation,
\be
    F_{\hat{\d}}[\hat{A}] = \star \, F_{\hat{\d}}[\hat{A}]    \label{sdYM-gen}
\ee
which is equivalent to
\be
   F_{\mu\nu} = 0 = F_{\bar{\mu}\bar{\nu}} \, , \qquad
   F_{\mu \bar{\rho}} \, \kappa^{\bar{\rho}}{}_\nu =  F_{\nu \bar{\rho}} \, \kappa^{\bar{\rho}}{}_\mu  \; .
\ee

Let $(\O(\A),\d)$ be the differential calculus over $\A$ which is obtained from the ordinary differential calculus by  regarding the coordinates $x^{\bar \mu}$ as parameters:
\be
    \d f = (\partial_\mu f) \, \d x^\mu    \; .
\ee
Correspondingly, let $(\bar{\O}(\A), \bar{\d})$ be the calculus obtained from the ordinary one by regarding the coordinates $x^\mu$ as parameters:
\be
    \bar{\d} f = (\partial_{\bar{\mu}} f) \, \bar{\d} x^{\bar{\mu}}   \; .
\ee
Then $\hat{\O}(\A)$ is the skew tensor product of $\O(\A)$ and $\bar{\O}(\A)$, and 
\be
    \hat{\d} = \d + \bar{\d}  \; .
\ee
 Furthermore, we define
\be
    \delta x^{\bar \mu}  = \kappa^{\bar{\mu}}{}_\nu \, \d x^\nu
                         = \kappa^{-1} ( \hat{\d} x^{\bar{\mu}} )
\ee
and
\be
     B = B_{\bar{\mu}} \, \delta x^{\bar{\mu}}
       = \kappa^{-1} (B_{\bar{\mu}} \, \hat{\d} x^{\bar{\mu}})  \; .
\ee
Now (\ref{sdYM-gen}) is found to be equivalent to
\be
     F_\d[A] = 0 = F_\delta[B] \, , \qquad
    \d B + \delta A + B \, A + A \, B =0 
\ee
which are the conditions (\ref{D2A})-(\ref{DADB}). By a gauge transformation, we can achieve that $B=0$ and thus $D_\delta = \delta$. Since $H^1_\delta(\O(\A))$ is trivial, the iterative construction of $\delta$-closed 1-forms in section 4 works. As a special case we recover the self-dual Yang-Mills equation in four real dimensions, see below. We have generalized this example to a set of integrable equations in $2n$ dimensions. Equations of this kind have also been considered in \cite{Ward84}.
\vskip.2cm
\noindent
{\em Example.} Let $\A = C^\infty(\mathbb{C}^2)$. In terms of complex coordinates $y,z$ with complex conjugates $\bar{y},\bar{z}$, we  introduce a bi-differential calculus via 
\be
     \delta f = f_{\bar{y}} \, \delta \bar{y} + f_{\bar{z}} \, \delta \bar{z}  \, , \qquad
     \d f = f_y \, \delta \bar{z} - f_z \, \delta \bar{y} \; .
\ee
With $A=g^{-1} \d g$ we have $F_\d[A] = 0$, and $\delta A = 0$ takes the form
\be
      (g^{-1} g_y)_{\bar{y}} + (g^{-1} g_z)_{\bar{z}} = 0
\ee
which is known to be equivalent to the self-dual Yang-Mills equation \cite{Yang77}. Indeed, in this case the map $\star$ defined above coincides with the Euclidean Hodge operator. The construction of conservation laws in the form given in \cite{BIZZ79} was carried over to the self-dual Yang-Mills equation in \cite{PSC79} and is easily recovered in our framework (see also \cite{Pohl80} for a different approach).
\hfill                $\diamondsuit$

\section{Conclusions}
\setcounter{equation}{0}
We have shown that, under certain conditions, a gauged bi-differential calculus (which has two flat covariant derivatives) leads to an infinite set of covariantly constant 1-forms. In many integrable models, these are realized by conserved currents, as we have demonstrated in particular for (principal) chiral models, some Toda models, the KP equation and the self-dual Yang-Mills equation. Other models are obtained via (dimensional) reduction of bi-differential calculi. For example, the KdV equation is a reduction of the KP equation and there is a corresponding reduction of the gauged bi-differential calculus which we associated with the KP equation. Many more examples are expected to fit into this scheme. Moreover, the latter leads to  possibilities of constructing new integrable models. In particular, the method is not restricted to certain (low) dimensions, as we have demonstrated in section \ref{sec:Plebanski} and section \ref{sec:sdYM equations}. 
We have also indicated the possibility of infinite sets of covariantly constant $s$-forms with $s > 1$, which still has to be explored.
\vskip.2cm

The question remains how our approach is related to various other characterizations of completely integrable systems. If a system with a Lax pair is given, this defines an operator $D_\d$. The problem is then to find another linearly independent operator $D_\delta$ such that $D_\delta^2 = 0 = D_\d D_\delta + D_\delta D_\d$. The existence of such a $D_\delta$ is not guaranteed, however, and may depend on the choice of Lax pair.
\vskip.2cm

Over the years several deep insights into soliton equations and integrable models have been achieved, in particular the AKNS scheme \cite{AKNS}, the $r$-matrix \cite{Fadd+Takh87} and the bi-Hamiltonian formalism 
\cite{Olve86}, Hirota's method \cite{Hiro76}, Sato's theory \cite{Sato81}, and relations with infinite dimensional Lie algebras \cite{Jimb+Miwa83}. 
To this collection of powerful approaches to the understanding and classification of soliton equations and integrable models, our work adds a new one which is technically quite simple and which is directly related to the physically important concept of conserved currents and charges. Besides the further clarification of relations with the approaches just mentioned, a generalization of the scheme presented in this work to supersymmetric models should be of interest.

\end{document}